\documentclass[%
 reprint,
 amsmath,amssymb,
 aip,
]{revtex4-2}

\usepackage{url}
\usepackage{booktabs}
\usepackage[capitalise]{cleveref}
\usepackage{graphicx}
\usepackage{dcolumn}
\usepackage{bm}

\begin{document}

\preprint{APS/123-QED}

\title{Nonlinear Response Relations and Fluctuation-Response Inequalities for Nonequilibrium Stochastic Systems}

\author{Jiming Zheng}
\email{jiming@unc.edu}
\affiliation{Department of Chemistry, University of North Carolina-Chapel Hill, NC}
\author{Zhiyue Lu}
\email{zhiyuelu@unc.edu}
\affiliation{Department of Chemistry, University of North Carolina-Chapel Hill, NC}

\date{\today}

\begin{abstract}
Predicting how systems respond to external perturbations far from equilibrium remains a fundamental challenge across physics, chemistry, and biology. We present a unified response framework for stochastic Markov dynamics that integrates linear and nonlinear perturbations. Our formalism expresses nonlinear responses of observables in terms of the covariance between the observable and a nonlinear conjugate variable. The nonlinear conjugate variable is subject to the complete Bell polynomial form and is determined by the stochastic entropy production. In addition, the Fluctuation-Response Inequalities (FRIs) are also derived for nonlinear responses, unraveling the general trade-off relations between nonlinear response and systems' fluctuations far from equilibrium. The validity of our theory is verified by the numerical results from a symmetric exclusion process (SEP). By unifying and extending nonequilibrium linear response theories, our approach can provide principled design rules for sensitive, adaptive synthetic and biological networks.

\end{abstract}

\maketitle

\section{Introduction}

Understanding how thermal systems respond to external parameters is a central goal in statistical physics, with fundamental implications across physics, chemistry, and biology. At thermal equilibrium, the system's response to perturbations is elegantly described by the equilibrium Boltzmann distribution. Dynamical perturbations near equilibrium are analyzed using the Green–Kubo fluctuation–dissipation theorem, which relates linear response functions directly to equilibrium correlation functions \cite{kubo1957statistical,kubo1966fluctuation}. However, many critical processes - such as nonequilibrium transport, protein folding, molecular binding, and chemical reactions - occur far from equilibrium, and are often transiently driven by continuously changing environments or intrinsic biochemical dynamics. For these processes, a general response theory that extends beyond equilibrium, and even far away from nonequilibrium steady states (NESS), is required to reliably predict and control their dynamic behaviors. Furthermore, to reveal the system's response toward external control beyond the linear regime, higher-order response sensitivity and even global (non-perturbative)response theory become useful. 

Historically, response theories near equilibrium predominantly employ Green's function approaches. Originated by Green and extensively developed by Kubo \cite{kubo1957statistical,kubo1966fluctuation}, this method computes linear response functions from correlation functions averaged over equilibrium-state distributions (i.e., reference ensembles). Nonlinear response relies on more complex correlation terms in the response function to account for the mismatch between the perturbed dynamics and the reference equilibrium dynamics. Extensions of Green's function methods to near-NESS conditions were pioneered by Agarwal \cite{agarwal1972fluctuation} and developed by others \cite{lippiello2008nonlinear,lippiello2008nonlinear2,wu2020generalized,zhang2021quantum,feng2011potential,marconi2008fluctuation}, although these formulations introduce additional complexity due to the breaking of detailed balance. In these approaches, the reference ensemble is the NESS state distribution of the unperturbed system. Nonlinear generalizations of response relations within Green’s function regime further address strong perturbations that drive system distributions prominently away from equilibrium or NESS, necessitating intricate multi-time correlation functions to correct for substantial mismatches between perturbed and reference ensembles \cite{diezemann2012nonlinear,kryvohuz2014nonlinear,mukamel1996classical,peterson1967formal}.

More recently, trajectory-based approaches have emerged, deriving response relations by explicitly comparing perturbed and unperturbed system trajectories. In particular, Seifert and Speck generalized the Agarwal form of linear response theory to NESS under the view of stochastic thermodynamics, systematically addressing the role of trajectory entropy in the linear response region around NESS \cite{seifert2010fluctuation}. Baiesi, Maes, and Wynants further provided a general linear response theory using trajectory ensembles, explicitly decomposing response functions into entropy-production and dynamical-activity (``frenetic'') contributions \cite{baiesi2009nonequilibrium,baiesi2013update,maes2020response}.
In recent years, significant progress has been made on nonequilibrium response theories, including a series of fluctuation-response inequalities (FRIs) \cite{zheng2025universal,ptaszynski2024dissipation,aslyamov2025nonequilibrium,ptaszynski2024nonequilibrium,ptaszynski2025nonequilibrium,kwon2024fluctuation,zheng2025spatial,liu2025dynamical}, graph theory analysis for Markov jump processes and chemical reaction networks \cite{owen2020universal,fernandes2023topologically,chun2023trade,owen2023size,harunari2024mutual}, and linear algebra analysis for Markov jump networks \cite{aslyamov2024general,aslyamov2024nonequilibrium,harunari2024mutual}.
However, existing nonequilibrium response theories are primarily focusing on linear perturbations or inequalities, and very few studies on nonlinear responses \cite{basu2015frenetic,hasegawa2019uncertainty,dechant2020fluctuation,harunari2024mutual,bao2024nonequilibrium}. A truly general trajectory framework capable of addressing nonlinear and non-perturbative responses of transient, far-from-equilibrium (non-NESS) processes has remained elusive.

In this work, we introduce a universal trajectory-level response theory valid for nonlinear and non-perturbative response phenomena in stochastic Markov systems arbitrarily far from equilibrium, including transient (non-NESS) regimes. The key insight is to recognize higher-order derivatives of the stochastic entropy along a single trajectory as the conjugate observables for nonlinear response. Applying the Cauchy--Schwarz inequality to the resulting identities then yields Fluctuation-Response Inequalities (FRIs) for arbitrary-order nonlinear response, which agree with existing linear FRIs \cite{ptaszynski2024dissipation,aslyamov2025nonequilibrium,ptaszynski2024nonequilibrium,ptaszynski2025nonequilibrium,liu2025dynamical,kwon2024fluctuation,zheng2025universal,zheng2025spatial} when restricted to $n=1$.

Compared with existing approaches, our framework offers three distinguishing features. First, it goes beyond linear response to \emph{arbitrary order} $n\ge 2$, and delivers exact closed-form estimators $\partial_\lambda^n\langle Q\rangle = \operatorname{Cov}(Q,B_n)$ that require only a \emph{single unperturbed simulation}, eliminating both the $N{+}1$ ensembles required by naive finite differencing in $N$-dimensional parameter space and the delicate choice of the perturbation step size $\Delta\lambda$. Second, it establishes a diagonal (uncorrelated) structure of the nonlinear conjugate variables $\{B_{n,ij}\}$ across transition edges, which allows response uncertainty relations to factorize at all orders. Third, it applies to transient far-from-equilibrium dynamics, not only to steady states. To the best of our knowledge, no previously existing framework provides closed-form trajectory estimators for arbitrary-order nonlinear responses in this generality.

We emphasize that our theory does not utilize the Green's function approach. As a result, different from existing Green's function nonlinear response theories \cite{lippiello2008nonlinear,lippiello2008nonlinear2,diezemann2012nonlinear,kryvohuz2014nonlinear,mukamel1996classical}, the nonlinear (high-order) response in our framework does not involve multi-time correction terms. Conceptually, trajectory ensemble approaches explicitly express probabilities of individual trajectories, directly contrasting with Green’s function methods. Green's function formalisms implicitly integrate over trajectories through propagators, hiding trajectory-level details behind convolutions and state-to-state probabilities. By explicitly highlighting trajectory probabilities, our framework not only clarifies the fundamental statistical structure of nonequilibrium response but also significantly simplifies practical computations.

\section{From Linear Response to Nonlinear Response}
We consider a general Markov system that describes the stochastic process of a particle. This particle could represent a physical particle or the microscopic state of a molecule in its state space. Its evolution can be characterized by the probability distribution $p(x(t))$ of the state or an ensemble of points $x(t)$ at various times. Alternatively, the system's evolution can be fully captured by stochastic trajectories $X_\tau = \{ x(t) \}_{t \in [0, \tau]}$ during the interval $[0, \tau]$. Here, the distribution of positions (or states) $p(x(t))$ can be recovered from the distribution of stochastic trajectories $\mathcal{P}[X_\tau]$. We further specify an observable $Q[X_\tau]$ as a functional of trajectories. The observable $Q[X_\tau]$ could either be a function of particle positions or transition events. Such observables are fundamental to many physics models and experiments, including diffusion \cite{levy1940certains,godreche2001statistics,majumdar2002local,lapolla2019manifestations,hartich2023violation}, active matter \cite{ramaswamy2010mechanics,ramaswamy2017active,marchetti2013hydrodynamics}, optics \cite{o2012time,margolin2005nonergodicity,ramesh2024arcsine,gopich2012theory}, chemical sensing \cite{bialek2005physical,endres2008accuracy,mora2010limits,govern2012fundamental,govern2014energy,harvey2023universal}, or biological transport \cite{maffeo2012modeling,catterall2010ion,roux2004theoretical}. The average of such an observable is given by a trajectory-ensemble average:
\begin{equation}
    \langle Q \rangle = \int\mathcal{D}[X_\tau] \mathcal{P}[X_\tau]Q[X_\tau].
    \label{eq: define Q}
\end{equation}
Assuming that there is a general parameter $\lambda$ controlling the dynamics of the system, such as the temperature of the thermal bath, the driving force applied to the system, or the internal energy barrier between different positions. The parameter $\lambda$ generally changes the distribution $p(x(t))$ as well as $\mathcal{P}[X_\tau]$. When the parameter is changed, the observable $\langle Q \rangle$ will deviate from its original value, which gives the response of the system represented on $Q$.

\subsection{Linear Response}
The linear response of a system's observable $Q$ is generally defined by the first-order derivative, $\partial_\lambda \langle Q \rangle$. The linear response of trajectory observables in stochastic Markov systems has been systematically developed in the trajectory-ensemble framework by Seifert and Speck~\cite{seifert2010fluctuation} and by Baiesi, Maes, and Wynants~\cite{baiesi2009nonequilibrium,baiesi2013update,maes2020response}. Here, we offer a slightly modified version to enhance its applicability and use it to derive higher-order responses. The linear response can be determined from understanding how trajectory probabilities change with the control parameter $\lambda$. This change is characterized by
\begin{equation}
    \frac{\partial \mathcal{P}[X_\tau; \lambda]}{\partial \lambda} = \Lambda[X_\tau; \lambda] \mathcal{P}[X_\tau; \lambda].
    \label{eq: trajectory prob response to lambda}
\end{equation}
Here $\Lambda [X_\tau,\lambda] \equiv \frac{\partial  \ln \mathcal{P}[X_{\tau} ; \lambda ]} {\partial  \lambda}$ is the score function.

Due to the normalization of trajectory probabilities, the expectation value of $\Lambda$ must always be zero. Using this property, \cref{eq: define Q,eq: trajectory prob response to lambda}, and the chain rule, a general response relation for any trajectory observable $\langle Q \rangle$ can be written as:
\begin{equation}
    \partial_\lambda \langle Q \rangle = \left\langle \frac{\partial Q}{\partial \lambda} \right\rangle + \operatorname{Cov}(Q, \Lambda), \label{eq: linear response relation}
\end{equation}
where the covariance function $\operatorname{Cov}(Q, \Lambda) \equiv \langle Q\Lambda \rangle - \langle Q \rangle \langle \Lambda \rangle = \langle Q\Lambda \rangle$ since $\langle \Lambda \rangle = 0$. 
In previous linear response studies \cite{seifert2010fluctuation,baiesi2009nonequilibrium,baiesi2013update,maes2020response}, the first term on the r.h.s of \cref{eq: linear response relation} is typically ignored due to the assumption that the observable $Q[X_{\tau}]$ does not explicitly depend on the control parameter $\lambda$. However, we emphasize here that it plays a crucial role in determining higher-order responses.

\subsection{Nonlinear Response}

The nonlinear response represented by an observable $Q$ is generally defined by the higher-order derivatives $\partial_\lambda^n \langle Q \rangle \equiv \partial^n\langle Q \rangle/\partial \lambda^n$, where $n\in \mathbb{Z}^+$ is a positive integer. In the following, we consider $\lambda$-independent observables for clarity. The more complicated $\lambda$-dependent case follows a chain rule expansion. The second-order response can be treated as the first-order response of the special observable ``$Q\Lambda$'':
\begin{subequations}
\begin{align}
    \partial^2_\lambda \langle Q\rangle &= \partial_\lambda \langle Q\Lambda \rangle \\
    &= \left\langle Q\frac{\partial\Lambda}{\partial\lambda}\right\rangle + \operatorname{Cov}(Q\Lambda, \Lambda)\\
    &= \left\langle Q \left( \frac{\partial \Lambda}{\partial \lambda} + \Lambda^2 \right)\right\rangle.
    \label{eq: 2nd-order response}
\end{align}
\end{subequations}
Here, the first term $\left\langle Q\frac{\partial\Lambda}{\partial\lambda}\right\rangle$ cannot be ignored since the new observable ``$Q\Lambda$'' is $\lambda$-dependent. Following this idea, we can calculate the $n$th-order response by taking derivatives on \cref{eq: linear response relation} $n-1$ times.

The iterative process raises a recurrence relation. For an observable $Q$ that does not explicitly depend on $\lambda$, the higher-order response is generally described by the higher-order derivatives of the trajectory probability $\mathcal{P}[X_\tau]$:
\begin{equation}
    \partial^n_\lambda \langle Q \rangle = \int\mathcal{D}[X_\tau] Q[X_\tau] \frac{\partial^n \mathcal{P}[X_\tau; \lambda]}{\partial \lambda^n}.
\end{equation}
For simplicity, we define a set of quantities $\{B_n\}$:
\begin{equation}
    B_n \equiv \frac{1}{\mathcal{P}[X_\tau; \lambda]} \frac{\partial^n \mathcal{P}[X_\tau; \lambda]}{\partial \lambda^n}.
    \label{eq: define Bn}
\end{equation}
which serves as the $n$-th order conjugate variable for higher-order response relations. As expected, $B_n$ reduces to the score function $\Lambda$ when $n = 1$ and reduces to $1$ when $n = 0$. Therefore, we obtain our first main result, the $n$th-order response given by the covariance between the observable $Q$ and $B_n$ as
\begin{equation}
    \partial^n_\lambda\langle Q \rangle = \langle Q B_n \rangle = \operatorname{Cov}(Q, B_n),
    \label{eq: nonlinear response relation}
\end{equation}
where we take advantage of $\langle B_n \rangle = 0$, which can be straightforwardly obtained by summing over all trajectories on both sides of \cref{eq: define Bn}. We noticed that $B_n$ is subject to the following recurrence relation:
\begin{equation}
\begin{cases}
    B_0 = 1, \\
    B_{n+1} = \Lambda B_n + \partial_\lambda B_n, & \quad n \ge 0 
    \label{eq: recurrence relation}
\end{cases}
\end{equation}
This relation indicates that $\{B_n\}$ are complete Bell polynomials \cite{bell1934exponential}. The general formula for $B_n$ is
\begin{equation}
    B_n(\Lambda, \cdots, \partial^{n-1}_\lambda\Lambda) = \sum_{\text{condition}} n! \prod_{i=1}^n \frac{1}{k_i!}\left(\frac{\partial^{i-1}_\lambda\Lambda}{i!}\right)^{k_i},
\end{equation}
where the summation condition is over all combinations of nonnegative integer $k_i$'s restricted by $k_1 + 2k_2 + \cdots + nk_n = n$. The appearance of the complete Bell polynomials here is a direct combinatorial consequence of differentiating the log-probability $\ln P[X_\tau;\lambda]$ via Fa\`a di Bruno's formula, analogous to the way Bell polynomials relate moments and cumulants.

The expressions for $B_n$'s are complicated, especially when $n$ is large. Here we list the first four in \cref{tab: Bn} for reference. As a verification, the expression of $B_2$ matches \cref{eq: 2nd-order response}.
\begin{table}[htbp]
    \centering
    \caption{Explicit expressions for $B_1$, $B_2$, $B_3$, and $B_4$.}
    \begin{tabular}{cl}
        $B_n$ & Explicit expression \\
        \midrule
        $B_1$ & $\Lambda$ \\
        $B_2$ & $\Lambda^2 + \partial_\lambda\Lambda$ \\
        $B_3$ & $\Lambda^3 + 3\Lambda\partial_\lambda\Lambda + \partial^2_\lambda\Lambda$ \\
        $B_4$ & $\Lambda^4 + 6\Lambda^2\partial_\lambda\Lambda + 4\Lambda\partial^2_\lambda\Lambda + 3(\partial_\lambda\Lambda)^2 + \partial^3_\lambda\Lambda$ \\
        \bottomrule
    \end{tabular}
    \label{tab: Bn}
\end{table}

\section{Properties of Response Conjugates}

\subsection{Physical Interpretations} 
We first discuss the physical interpretations of the response conjugate variable $B_n$. In stochastic thermodynamics, the logarithm of the trajectory probability $\ln\mathcal{P}[X_\tau]$ is specified as the trajectory entropy production $\sigma[X_\tau]$ of the system \cite{seifert2012stochastic,peliti2021stochastic}. As stated by Seifert and Speck \cite{seifert2010fluctuation}, the linear conjugate variable $\Lambda$ can be understood as the susceptibility of the stochastic entropy $\partial_\lambda \sigma(\lambda)$. Here, for the nonlinear conjugate variable $B_n$, it is a function of higher-order derivatives of $\Lambda$ as $B_n \equiv B_n(\Lambda, \cdots, \partial^{n-1}_\lambda\Lambda)$. Therefore, $B_n$ is entirely determined by all the $n$th-order derivatives of the stochastic entropy $\sigma$ as $B_n \equiv B_n(\partial_\lambda \sigma, \cdots, \partial^n_\lambda \sigma)$. The trajectory entropy production can be written as the difference between the total entropy production $\sigma_{\text{tot}}$ and the entropy production of the reservoir $\sigma_{\text{res}}$. Therefore, the contribution of the conjugate quantity $B_n$ can be written as the convolution of two parts:
\begin{equation}
    B_n(\sigma) = \sum_{k=0}^n \frac{n!}{k!(n-k)!} B_k(\sigma_{\text{tot}})B_{n-k}(-\sigma_{\text{res}}),
    \label{eq: tot - res}
\end{equation}
where we use $B_n(\sigma) \equiv B_n(\partial_\lambda \sigma, \cdots, \partial^n_\lambda \sigma)$ for the abbreviation. For the linear case, i.e., $n = 1$, \cref{eq: tot - res} simplifies to $\partial_\lambda \sigma = \partial_\lambda \sigma_\text{tot} - \partial_\lambda \sigma_{\text{res}}$ \cite{seifert2010fluctuation}. However, for higher-order cases, $\{B_n\}$ are not separable. The contributions of $\sigma_{\text{tot}}$ and $\sigma_{\text{res}}$ must be written as \cref{eq: tot - res}. For equilibrium systems, the total entropy production $\sigma_{\text{tot}}$ is zero, and the system's $n$th-order response is solely dependent on $B_n(-\sigma_{\text{res}})$.

Despite the decomposition $\sigma_{\text{tot}} = \sigma + \sigma_{\text{res}}$, the total entropy production $\sigma_{\text{tot}}$ can also be decomposed into two parts, namely the housekeeping part $\sigma_{\text{hk}}$ and the excess part $\sigma_{\text{ex}}$ \cite{seifert2012stochastic,peliti2021stochastic}. There exist different methods for the decomposition, such as the Hatano-Sasa decomposition \cite{hatano2001steady} and the Maes-Neto\v{c}n\`{y} decomposition \cite{maes2014nonequilibrium}. For systems in nonequilibrium stead states, all the types of decomposition give $\sigma_{\text{hk}} = 0$ \cite{dechant2022geometric}. Therefore, the steady-state response can be written as
\begin{equation}
    B_n(\sigma) = \sum_{k=0}^n \frac{n!}{k!(n-k)!} B_k(\sigma_{\text{ex}})B_{n-k}(-\sigma_{\text{res}}).
    \label{eq: ex - res}
\end{equation}
For systems without nonconservative forces, the excess part $\sigma_{\text{ex}}$ is zero \cite{dechant2022geometric}. So the response for such systems can be written as
\begin{equation}
    B_n(\sigma) = \sum_{k=0}^n \frac{n!}{k!(n-k)!} B_k(\sigma_{\text{hk}})B_{n-k}(-\sigma_{\text{res}}).
    \label{eq: hk - res}
\end{equation}

\subsection{Explicit Forms and Statistical Properties}
The above analysis for linear and nonlinear response generally holds for all types of stochastic dynamics. Here, we take discrete Markov jump dynamics to illustrate the explicit expressions and statistical properties of $B_n$.

Consider the $m$-state Markov jump process controlled by the master equation:
\begin{equation}
    \frac{\partial\boldsymbol{p}(t)}{\partial t} = R \boldsymbol{p}(t),
    \label{eq: master equation}
\end{equation}
where $\boldsymbol{p}(t) = (p_1(t), p_2(t), \cdots, p_m(t))^\top$ is the vector of probabilities on the $m$ states and $R = \{R_{ij}\}_{m\times m}$ is the transition rate matrix with diagonal elements defined as $R_{ii} = -\sum_j R_{ji}$. Each realization of the system's evolution within a time duration can be specified by a series of dwell periods on states and instantaneous jump events between states. That gives the stochastic trajectory for Markov jump processes:
\begin{equation}
    X_\tau = ((x_0, t_0), (x_1, t_1), \cdots, (x_\alpha, t_\alpha), \cdots, (x_N, t_N), \tau).
\end{equation}
where $x_0$ is the initial state, $t_0 = 0$ is the initial time, $t_\alpha$ is the time of the $\alpha$-th jump event from the previous state to the new state $x_\alpha$, and $\tau$ is the total time length of the trajectory.
The products of dwelling probability and jump probability give the trajectory probability:
\begin{subequations}
\begin{align}
    \mathcal{P}[X_\tau] &= p_{x_0}(t_0) \prod_{\alpha=1}^{N} R_{x_\alpha x_{\alpha-1}} \prod_{\alpha=0}^{N} e^{\int_{t_\alpha}^{t_{\alpha+1}} R_{x_\alpha x_\alpha}\mathrm{d}t} \\
    &= p_{x_0}(t_0) \prod_{i \neq j} e^{N_{ij}\ln R_{ij} - R_{ij}T_j},
    \label{eq: trajectory probability}
\end{align}
\end{subequations}
where $N_{ij}[X_\tau]$ is the number of transitions from state $j$ to $i$ and $T_j[X_\tau]$ is the total dwelling time on the state $j$ in the given trajectory $X_\tau$. In the second equation, we group the same types of transitions and dwellings and use $R_{ii} = -\sum_j R_{ji}$.

\subsubsection{Linear Conjugate Variable} In general, a control parameter $\lambda$ can impact the value of one or many transition rates $\{R_{ij}\}$. From here and below, we use $B_{n, ij}$ to represent the corresponding conjugate variables when choosing $\lambda = \ln R_{ij}$. By doing so, the linear conjugate variable $\Lambda$ becomes the {\it dynamical discrepancy} defined for the corresponding edge $j\rightarrow i$: 
\begin{equation}
    \Lambda_{ij}[X_\tau] \equiv N_{ij}[X_\tau] - R_{ij}T_{j}[X_\tau].
    \label{eq: Lambda_ij define}
\end{equation}
For each realization of the stochastic process $X_{\tau}$, $\Lambda_{ij}$ quantifies the stochastic mismatch between the number of jumps $N_{ij}$ and the transition-rate-weighted dwelling time $R_{ij}T_j$. We acknowledge that Seifert and Speck obtained a similar expression for $\Lambda$ \cite{seifert2010fluctuation}. We also notice that the short-time dynamical discrepancy is recognized as noise for the master equation in the recent study \cite{torbjorn2025stochastic}. For a general control parameter (impacting multiple transition rates), the conjugate variable $\Lambda$ can be decomposed into a weighted summation of the dynamical discrepancies $\{\Lambda_{ij}\}$ on all $\lambda$-controlled transition edges:
\begin{equation}
    \Lambda = \sum_{i\neq j} \frac{\partial  \ln R_{ij}}{\partial  \lambda} \Lambda_{ij}.
    \label{eq: Lambda decomposition}
\end{equation}
For systems governed by time-dependent transition rates, this expression can be slightly modified into a time integral.

The dynamical discrepancy $\{ \Lambda_{ij} \}$ has interesting statistical properties. Firstly, they each satisfy the zero-mean property as a consequence of $\langle N_{ij} \rangle = R_{ij}\langle T_j \rangle$. Secondly, the dynamical discrepancies of different transition edges $\{ \Lambda_{ij} \}$ are statistically independent of each other. Moreover, the edge-edge correlations of $\{ \Lambda_{ij} \}$ between any pair of transition edges are:
\begin{equation}
    \operatorname{Cov}(\Lambda_{ij}, \Lambda_{kl}) = \begin{cases}
        \langle N_{ij} \rangle, & i = k \text{ and } j = l, \\
        0, & \text{else}.
    \end{cases}
    \label{eq: spatial correlation of Lambda_ij}
\end{equation}
This relation can be proven by taking $Q = \Lambda_{ij}$ and $\lambda = \ln R_{kl}$ in \cref{eq: linear response relation}. It indicates that the conjugate variables for localized control parameters applied to different edges are statistically independent. In addition, each edge's dynamical discrepancy variance characterizes the averaged transition number on the corresponding edge. As a result, the covariance matrix of $\{ \Lambda_{ij} \}$ is diagonal.

Since $\Lambda$ is a linear combination of edge-wise $\{\Lambda_{ij}\}$, it inherits properties from $\{\Lambda_{ij}\}$. The variance of $\Lambda$ is also known as the Fisher information $\mathcal{I}(\lambda)$. It turns out to be a linear combination of averaged transition numbers:
\begin{subequations}
\begin{align}
    \mathcal{I}(\lambda) &= \operatorname{Var}[\Lambda] = -\left\langle \frac{\partial \Lambda}{\partial \lambda} \right\rangle
    \label{eq: variance Lambda} \\
    &= \sum_{i\neq j} \left( \frac{\partial \ln R_{ij}}{\partial \lambda} \right)^2 \langle N_{ij} \rangle
    \label{eq: trajectory FI}.
\end{align}
\end{subequations}
Again, the variance of $\Lambda$ for a time-inhomogeneous process should be written as a time integral. By choosing $\lambda=R_{ij}$, it reduces to the result in \cite{zheng2025spatial}: $\mathcal{I} = \langle N_{ij} \rangle / R_{ij}^2$. Our theory also refines the result in \cite{dechant2020fluctuation} with Fisher information $\mathcal{I}(\lambda) = \langle Z_{ij}^2 R_{ij} T_j \rangle$ given the transition rates following $R_{ij} = k_{ij}e^{\lambda Z_{ij}}$.

\subsubsection{Nonlinear Conjugate Variables}
The nonlinear conjugate variable $B_n$ shares similar properties with $\Lambda$. First, by summing over all trajectories on both sides of \cref{eq: define Bn}, we find that $\langle B_n \rangle = 0$. Then, by taking trajectory-ensemble average on both sides of the recurrence relation \cref{eq: recurrence relation}, we immediately get $\langle B_{n+1} \rangle = \langle \Lambda B_n \rangle + \langle \partial_\lambda B_n \rangle = 0$, which leads to $\langle \Lambda B_n \rangle = -\langle \partial_\lambda B_n \rangle$. In this case, the calculation for the trajectory Fisher information in \cref{eq: variance Lambda} is a special case for $n = 1$.

When the system follows the master equation \cref{eq: master equation} and has a trajectory description as \cref{eq: trajectory probability}, the uncorrelated property \cref{eq: spatial correlation of Lambda_ij} also applies to $B_n$. Since $B_n$ is a function of $\{ \Lambda, \partial_\lambda\Lambda, \cdots, \partial^{n-1}_\lambda\Lambda \}$, the only parameter in $B_{n, ij}$ is $\ln R_{ij}$. Therefore, we obtain $\operatorname{Cov}(B_{n, ij}, B_{n, kl}) = 0$ when $i \neq k$ or $j \neq l$ (see Appendix for details). It indicates that, given the choice of $\lambda_{ij} = \ln R_{ij}$, the covariance matrix of $\{B_{n, ij}\}$ is diagonal for all $n$. However, in contrast to the linear conjugate variable, the edge decomposition property \cref{eq: Lambda decomposition} does not hold for $B_n$ since it is not a consequence of the uncorrelated property. Consider a general observable $Q$ that is not explicitly dependent on $\lambda$, the cross derivative $\frac{\partial^2 \langle Q \rangle}{\partial\ln R_{ij}\partial\ln R_{kl}} = \langle Q\Lambda_{ij}\Lambda_{kl} \rangle$ is generally not zero. Thus, the cross terms of different edges do not vanish for higher-order responses.

The variance of $B_n$ is characterized by $\operatorname{Var}[B_n] = \langle B_n^2 \rangle$. We find that
\begin{equation}
    \operatorname{Var}[B_n] = \langle B_n^2 \rangle = -\sum_{k=1}^n \frac{n!}{k!(n-k)!} \left\langle B_{n-k} \frac{\partial^k B_n}{\partial\lambda^k} \right\rangle.
    \label{eq: variance Bn}
\end{equation}
One can show that the Fisher information formula is a special case of \cref{eq: variance Bn}. When the system evolves under a master equation, the explicit expressions of $\operatorname{Var}[B_n]$ can be obtained. Specifically, by choosing $\lambda = \ln R_{ij}$, we have $\Lambda = N_{ij} - R_{ij}T_{j}$. In this case, all the higher-order derivatives of $\Lambda$ are $\partial^n_\lambda\Lambda = -R_{ij}T_j$. Therefore, the calculations for $B_n$ and its variance are greatly simplified. Specifically, we obtain
\begin{align}
    B_{2, ij} &= \Lambda_{ij}^2 - R_{ij} T_j, \\
    \operatorname{Var}[B_{2, ij}] &= \langle (2\Lambda_{ij}^2 + 3\Lambda_{ij} - 2R_{ij}T_j + 1) R_{ij}T_j \rangle
\end{align}
for $n = 2$.

Additionally, the complete Bell polynomial follows the weighted homogeneous property:
\begin{equation}
    B_n(c\partial_\lambda\sigma, c^2\partial^2_\lambda\sigma, \cdots, c^n\partial^n_\lambda\sigma) = c^n B_n(\partial_\lambda\sigma, \partial^2_\lambda\sigma, \cdots, \partial^n_\lambda\sigma),
\end{equation}
where $c \in \mathbb{R}$ is a constant. It reveals that if the amplitude of the control parameter is scaled by $c$ times, then the $n$th-order response is magnified by $c^n$ times. By choosing $c = -1$, i.e., reversal of the control parameter's direction, we find that the odd-order responses will be reversed while the even-order responses remain unchanged.

\subsubsection{Operational remarks on $\Lambda$ and $B_n$.} We emphasize that $\Lambda$ and $B_n$ are not independently postulated observables: they are fully determined by (i) the analytic dependence of the transition rates on the control parameter, $\partial_\lambda \ln R_{ij}$, which is known from the chosen parameterization, and (ii) the microscopic jump counts $N_{ij}[X_\tau]$ and dwell times $T_j[X_\tau]$, which are directly measurable from each trajectory. Given full access to the microscopic dynamics, $\Lambda$ and all $B_n$ are therefore computable without any perturbed simulation, finite-difference stepping, or propagator inversion. Our identities \cref{eq: nonlinear response relation} and \cref{eq: nonlinear FRIs} rely only on (a) differentiability of $P[X_\tau;\lambda]$ in $\lambda$ to the required order, so that $B_n$ is well defined, and (b) finiteness of the relevant moments $\langle Q^2\rangle$ and $\langle B_n^2\rangle$. At singular points of the parameterization, e.g., where a transition rate vanishes or diverges, or where the parameterization is non-smooth, $B_n$ and its variance may diverge and the bounds become uninformative (though still not incorrect). The predictions are directly testable by comparing $\operatorname{Cov}(Q,B_n)$ at a fixed $\lambda_0$ against finite-difference derivatives $\partial_\lambda^n\langle Q\rangle$ obtained from independent perturbed simulations, as demonstrated in Section~\ref{sec: numerics}.

\section{Nonlinear Fluctuation-Response Inequalities}
Recent studies have established the rich structure for FRIs \cite{zheng2025universal,zheng2025spatial,ptaszynski2024dissipation,aslyamov2025nonequilibrium,ptaszynski2024nonequilibrium,ptaszynski2025nonequilibrium,liu2025dynamical,kwon2024fluctuation}. Applying the Cauchy-Schwarz inequality $(\operatorname{Cov}(X, Y))^2 \le \operatorname{Var}[X] \operatorname{Var}[Y]$ to the linear response relation \cref{eq: linear response relation} yields the linear FRIs in \cite{zheng2025universal,zheng2025spatial,ptaszynski2024dissipation,aslyamov2025nonequilibrium,ptaszynski2024nonequilibrium,ptaszynski2025nonequilibrium,liu2025dynamical,kwon2024fluctuation}. The Cauchy-Schwarz inequality also applies to the nonlinear response relation \cref{eq: nonlinear response relation}. As a result, we obtain the second main result, nonlinear FRIs for $\lambda$-independent observables:
\begin{equation}
    \frac{(\partial^n_\lambda \langle Q \rangle)^2}{\operatorname{Var}[Q]} \le \operatorname{Var}[B_n].
    \label{eq: nonlinear FRIs}
\end{equation}
In the following, we illustrate this result for two general cases: one is a multi-dimensional parameter case, and the other is a choice of parameters to obtain Thermodynamic Uncertainty Relations (TURs) from the nonlinear response relations.

We first illustrate this result for the multi-dimensional parameter case. Recent studies show that the multi-dimensional Cram\'er-Rao inequality for linear response leads to the Response Uncertainty Relations (RURs) and further leads to the Response-TUR \cite{zheng2025spatial,ptaszynski2024dissipation,aslyamov2025nonequilibrium,ptaszynski2024nonequilibrium,ptaszynski2025nonequilibrium,kwon2024fluctuation}. In those cases, the Fisher information matrix is diagonal, so that the matrix form inequality becomes the RURs. Here, for nonlinear responses, the spatial correlation of $\{B_{n, ij}\}$ also implies that the multi-parameter covariance matrix of $\{B_{n, ij}\}$ is diagonal. Therefore, the same statement for the linear RURs \cite{zheng2025spatial,kwon2024fluctuation} also applies to the nonlinear ones. Considering the symmetric and asymmetric kinetic parameters
\begin{equation}
    R_{ij} = e^{b_{ij} + f_{ij} / 2},
\end{equation}
where $b_{ij} = b_{ji}$ represents the energy barrier and $f_{ij} = - f_{ji}$ represents the driving force on the edge. The following nonlinear RURs hold:
\begin{subequations}
\begin{align}
    \sum_{i<j} \frac{(\partial^n_{b_{ij}}\langle Q \rangle)^2}{\operatorname{Var}[B_{n, b_{ij}}]} &\le \operatorname{Var}[Q], \\
    \sum_{i<j} \frac{(\partial^n_{f_{ij}}\langle Q \rangle)^2}{\operatorname{Var}[B_{n, f_{ij}}]} &\le \operatorname{Var}[Q],
\end{align}
\end{subequations}
where $B_{n, b_{ij}}$ and $B_{n, f_{ij}}$ stands for the nonlinear conjugate variable for $b_{ij}$ and $f_{ij}$, respectively. For $n = 2$, the explicit form of $B_{2, b_{ij}}$ and $B_{2, f_{ij}}$ are:
\begin{equation}
    B_{2, *_{ij}} =(\Lambda_{*_{ij}})^2 - \partial_{*_{ij}}\Lambda_{*_{ij}}
\end{equation}
where $*$ is $f$ or $b$, $\Lambda_{b_{ij}} = \Lambda_{ij} + \Lambda_{ji}$, $\Lambda_{f_{ij}} = \frac{1}{2}\Lambda_{ij} - \frac{1}{2}\Lambda_{ji}$, $\partial^n_{b_{ij}}\Lambda_{b_{ij}} = -(R_{ij}T_{j} + R_{ji}T_i)$, and $\partial^n_{f_{ij}}\Lambda_{f_{ij}} = -\left(\frac{1}{2}\right)^{n+1} R_{ij}T_j - \left(-\frac{1}{2}\right)^{n+1} R_{ji}T_i$.

Here, we discuss the connection between \cref{eq: nonlinear FRIs} and TURs. It has been well studied that the linear FRI can be reduced to the TUR by choosing the parameterization as \cite{dechant2020fluctuation}
\begin{equation}
    R_{ij} = k_{ij} \exp\left( \lambda \cdot \frac{k_{ij}\pi_j - k_{ji}\pi_i}{k_{ij}\pi_j + k_{ji}\pi_i} \right).
    \label{eq: TUR parameter}
\end{equation}
Around $\lambda = 0$, it keeps the steady state distribution $\pi$ unchanged and the current response becomes the current itself, i.e., $\partial_\lambda\langle J \rangle = \langle J \rangle$ for any time-asymmetric current $J$. If one replace the parameter $\lambda$  in \cref{eq: TUR parameter} with $\frac{1}{n!}\lambda^n$, then the higher-order response is also the current itself, $\partial^n_\lambda \langle J \rangle = \langle J \rangle$. Therefore, we can obtain a series of uncertainty relations from the nonlinear FRI by using the higher order parameterization $\lambda \mapsto \frac{1}{n!}\lambda^n$:
\begin{equation}
    \frac{\langle J \rangle^2}{\operatorname{Var}[J]} \le \operatorname{Var}[B_n], \quad n \in \mathbb{Z}_+.
    \label{eq: extended TURs}
\end{equation}
Equality in \cref{eq: extended TURs} is attained when and only when the current $J$ and the nonlinear conjugate variable $B_n$ are linearly dependent as random variables on the trajectory space (up to an additive constant), as dictated by the saturation condition of the Cauchy--Schwarz inequality. For a generic time-antisymmetric current $J$ and the Bell-polynomial form of $B_n$, this exact linear dependence does not hold, so the bound \cref{eq: extended TURs} is generically not saturated. The gap between the two sides, therefore, provides a quantitative measure of how closely the current aligns, in the trajectory-ensemble sense, with the $n$th-order conjugate variable. In particular, different orders $n$ may yield different tightness, which motivates searching for an optimal $n$ for a given system. Under this parameterization, the case of $n=1$ reveals the TUR for Markov jump processes, i.e., $\operatorname{Var}[\Lambda]$ upper-bounds the precision by the pseudo-entropy production \cite{dechant2020fluctuation}. However, $\operatorname{Var}[B_n]$ is not guaranteed to be monotonic with respect to $n$ in general, as shown in the SI. Therefore, there may exist an optimal $n$ for \cref{eq: extended TURs} to get the tightest bound.

\section{Numerical Verifications}
\label{sec: numerics}

\begin{figure}[htbp]
\centering
\includegraphics[width=0.5\linewidth]{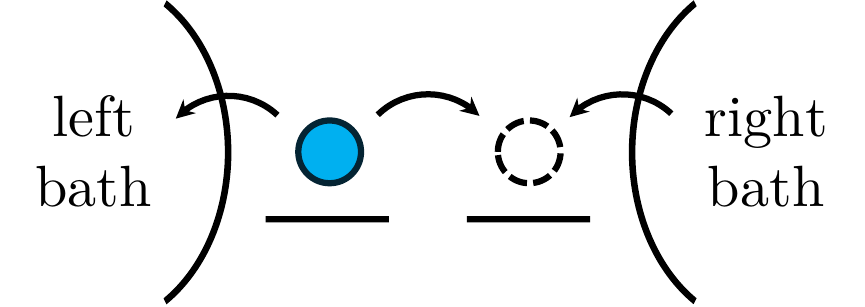}
\caption{Exclusion process (EP) model with two sites in the tube. Particles can hop between neighboring sites and between the bath and the contacting site. There are three possible transitions for the illustrated configuration, shown by curved arrows. The blue particle can jump to the left or the right, and a particle can jump from the right bath to the empty site.}
\label{fig: SEP model}
\end{figure}

To concretely illustrate the validity of our trajectory-based response framework, we consider a minimal symmetric exclusion process model representing particle transport through a narrow channel connecting two particle reservoirs with different concentrations, as shown in \cref{fig: SEP model}. Particles can hop symmetrically to neighboring sites with equal probability, subject to an exclusion rule: hops occur only if the target site is empty. Such symmetric exclusion processes, driven by concentration gradients imposed by boundary baths, are prototypical models for studying nonequilibrium transport phenomena. Transport between reservoirs has been extensively studied via two complementary routes. Mean-field and coarse-grained approaches reduce the problem to a small number of effective variables; while computationally efficient, they can miss correlations that are essential for sensing and energy-transduction accuracy, as highlighted, e.g., by Yuly, Zhang, and Beratan in the context of reversible electron bifurcation~\cite{terai2023correlated} and in our earlier work on biochemical transduction~\cite{slowey2022sloppy}. Detailed master-equation approaches, in contrast, retain full microscopic resolution.

\begin{figure}[htbp]
\centering
\includegraphics[width=1\linewidth]{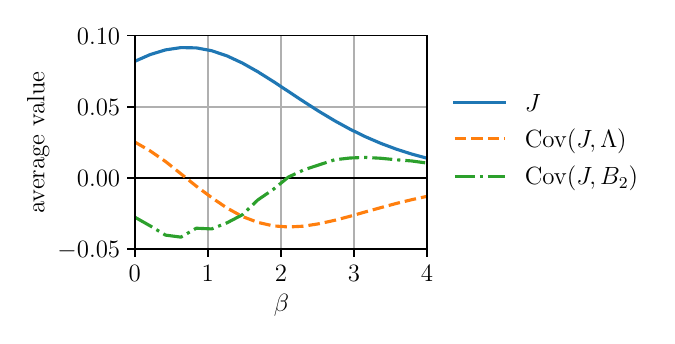}
\caption{Averaged currents, first-order response, and second-order response towards $\beta$. All three curves are sampled from a Monte-Carlo simulation with $10^8$ trajectories on a two-site exclusion process (EP) model. Each trajectory evolves up to $\tau = 1000$. The energy of a particle on any site is $1.0$, and the energy of a particle in either bath is $0.0$. All the energy barriers of particle hopping transitions (site-site hopping and site-bath hopping) are set to $1.0$. The concentration of the left bath is $2.5$, and is $5.0$ for the right bath.}
\label{fig: current and response}
\end{figure}

In this minimalistic model, we specifically investigate how temperature affects the nonequilibrium particle flow. To introduce an interesting and physically relevant nonmonotonic temperature dependence, we utilize particle-site repulsion to thermodynamically penalize particles in the sites (see energy details in \cref{fig: current and response}). As a result, the transport rate is both nonlinear and nonmonotonic with respect to environmental temperature, as shown in \cref{fig: current and response}. The particle-site repulsion is introduced so that the model exhibits a physically interesting \emph{nonmonotonic} temperature dependence of the current: at high $\beta$ the repulsion suppresses channel occupancy and the flow is quenched, while at low $\beta$ the channel saturates and jamming suppresses the flow, so an optimum emerges at intermediate $\beta$. This non-monotonicity is reminiscent of counterintuitive temperature dependence in phenomena such as the Mpemba effect and in transport through interacting channels \cite {schutz1997exact,liggett1975ergodic,giardina2006direct}. Importantly, for our purposes, it produces a control-parameter value ($\beta\approx 0.8$) at which the linear response identically vanishes, giving a clean testbed on which the second-order response predicted by our theory is the leading non-vanishing contribution.

Utilizing kinetic Monte Carlo simulations in conjunction with our trajectory-based response theory, we explicitly computed the first- and second-order derivatives of the particle flow rate with respect to the inverse temperature, $\beta$. The current-$\beta$ curve is maximized (flat) around $\beta = 0.8$, agreeing with our result of the first-order response curve crossing zero at that point. Furthermore, the inversion point of the current-$\beta$ curve around $\beta=2.0$ should correspond to a minimum of the first-order response curve, which is observed; also, the second-order response at the inversion point is indeed zero as in \cref{fig: current and response}. At $\beta\approx 0.8$ the first-order response $\partial_\beta\langle J\rangle$ identically vanishes, so linear response theory alone provides no information about how the current varies with $\beta$ near this point. Our framework nevertheless remains predictive: the second-order response $\operatorname{Cov}(J,B_2)$, shown in \cref{fig: current and response}, is the leading nonvanishing sensitivity and correctly identifies $\beta\approx 0.8$ as a local extremum of $\langle J\rangle(\beta)$. We emphasize that the quantities plotted in \cref{fig: current and response}, $\operatorname{Cov}(J,\Lambda)$ and $\operatorname{Cov}(J,B_2)$, are \emph{exact estimators} of $\partial_\beta\langle J\rangle$ and $\partial_\beta^2 \langle J\rangle$ via \cref{eq: nonlinear response relation}. 

In concrete terms, the example illustrates three features that our framework captures systematically: (i) the current-$\beta$ curve is nonmonotonic, with a maximum near $\beta\approx 0.8$ and an inflection near $\beta\approx 2.0$; (ii) the linear response $\operatorname{Cov}(J,\Lambda)$ vanishes at the maximum, and the second-order response $\operatorname{Cov}(J,B_2)$ vanishes at the inflection, both in quantitative agreement with the locations of the extrema of $\langle J\rangle(\beta)$; and (iii) all of these response functions are obtained from a single unperturbed simulation, without perturbed ensembles or step-size tuning. In contrast to mean-field transport theories, which would smear out these features, and to finite-difference master-equation approaches, which would require a separate simulation for each parameter value and each order, our trajectory-based estimators provide exact closed-form access to the full nonlinear response at the detailed master-equation level. This illustrates the utility of the framework for designing and controlling complex transport processes with many microscopic parameters.

\section{Conclusion}
In this paper, we propose a general trajectory-level framework of nonlinear response relations for stochastic processes arbitrarily far from equilibrium. Our first main result \cref{eq: nonlinear response relation} expresses the $n$th-order sensitivity of any trajectory observable as a single correlation between that observable and a nonlinear conjugate variable $B_n$. The recurrence relation of $B_n$ reveals a Bell-polynomial structure that unifies linear and nonlinear response. In our framework, the stochastic entropy production plays a key role in determining the nonlinear response, largely extending the statement in the linear response region \cite{seifert2010fluctuation}. Our second main result \cref{eq: nonlinear FRIs} generalizes the FRIs from linear response to arbitrary order nonlinear response. It shows that the trade-off relation between the response and internal fluctuations is general for nonequilibrium systems to arbitrary order perturbations. Furthermore, the correlation form of the response provides unbiased estimators at a fixed control parameter, eliminating the need for re-simulating perturbed dynamics and enabling efficient gradient and higher-order derivative evaluation. Our results provide a statistically estimable calculus for predicting and designing nonequilibrium responses, with immediate implications for optimizing control protocols and engineering targeted behavior in molecular, chemical, and active systems.

\begin{acknowledgments}
We appreciate funding support from the National Science Foundation under grant number DMR-2145256, and funding support from the Alfred P. Sloan Foundation Award under grant number G-2025-25194. We enjoy inspiring discussions with Prof. Chris Jarzynski, Prof. Hong Qian, Prof. Tim Elston, and Dr. Zhongmin Zhang.
\end{acknowledgments}

\appendix
\section{Covariance and Variance} \label{SIsec: covariance and variance}

In the following, we calculate the covariance and variance of $\{\Lambda_{ij}\}$ and $\{B_{n, ij}\}$.

\subsection{Linear Conjugate Variable} \label{SIsubsec: Lambda}

Let us take $Q = \Lambda_{ij}$, and consider the response $\partial_{\ln R_{kl}} \langle \Lambda_{ij} \rangle$. The response is given by
\begin{equation}
    \partial_{\ln R_{kl}} \langle \Lambda_{ij} \rangle = \langle \partial_{\ln R_{kl}} \Lambda_{ij} \rangle + \operatorname{Cov}(\Lambda_{ij}, \Lambda_{kl}) = 0.
\end{equation}
The response is always zero since $\langle\Lambda_{ij}\rangle = 0$ for any parameter. If $i = k$ and $j = l$, then $\Lambda_{ij} = N_{ij} - R_{ij} T_j$ depends on $\ln R_{ij}$. Therefore, we have
\begin{equation}
    \operatorname{Var}[\Lambda_{ij}] = \operatorname{Cov}(\Lambda_{ij}, \Lambda_{ij}) = -\langle \partial_{\ln R_{ij}} \Lambda_{ij} \rangle = \langle R_{ij} T_j \rangle = \langle N_{ij} \rangle.
\end{equation}
If $i \neq k$ or $j \neq l$, then $\Lambda_{ij} = N_{ij} - R_{ij} T_j$ is independent of $\ln R_{kl}$. Therefore, we have
\begin{equation}
    \operatorname{Cov}(\Lambda_{ij}, \Lambda_{kl}) = -\langle \partial_{\ln R_{kl}} \Lambda_{ij} \rangle = 0.
\end{equation}

Here, we offer another method to calculate $\operatorname{Var}[\Lambda_{ij}]$. The first-order derivative of $\mathcal{P}[X_\tau]$ defines the conjugate variable $\Lambda$:
\begin{equation}
    \frac{\partial \mathcal{P}}{\partial \lambda} = \Lambda \mathcal{P}.
\end{equation}
Now we consider its second-order derivative:
\begin{equation}
    \frac{\partial^2 \mathcal{P}}{\partial \lambda^2} = \frac{\partial \Lambda}{\partial \lambda} \mathcal{P} + \Lambda^2 \mathcal{P}.
\end{equation}
Integrating the equation over all the trajectories on both sides and assuming that the integral and the derivative are commutative. As a result, we have the following
\begin{equation}
    0 = \frac{\partial^2}{\partial\lambda^2}\int\mathcal{D}[X_\tau] \mathcal{P}[X_\tau] = \langle \partial_{\lambda}\Lambda\rangle + \langle \Lambda^2 \rangle.
\end{equation}
Therefore, the variance is $\operatorname{Var}[\Lambda_{ij}] = \langle \Lambda^2_{ij} \rangle = -\langle \partial_{\ln R_{ij}} \Lambda_{ij} \rangle = \langle R_{ij} T_j \rangle = \langle N_{ij} \rangle$.

\subsection{Non-linear Conjugate Variable} \label{SIsubsec: B_n}

Let us take $Q = B_{n, ij}$, and consider the response $\partial^n_{\ln R_{kl}} \langle B_{n, ij} \rangle$. If $i \neq k$ or $j \neq l$, then $B_{n, ij}$ is independent of $\ln R_{kl}$. In this case, we have
\begin{equation}
    \partial^n_{\ln R_{kl}} \langle B_{n, ij} \rangle = \operatorname{Cov}(B_{n, ij}, B_{n, kl}) = 0,
\end{equation}
where $\langle B_{n, ij} \rangle = 0$ for all parameters. If $i = k$ and $j = l$, then $B_{n, ij}$ depends on $\ln R_{ij}$. In this case, we need to use the chain rule expansion. For clarity, we use the second approach, which directly deals with trajectory probability. Consider the $2n$th-order derivative of the trajectory probability:
\begin{subequations}
\begin{align}
    \frac{\partial^{2n} \mathcal{P}}{\partial \lambda^{2n}}
    &= \frac{\partial^n (B_n\mathcal{P})}{\partial \lambda^n} \\
    &= \sum_{k = 0}^n \frac{n!}{k!(n-k)!} \frac{\partial^k B_{n}}{\partial \lambda^k} \frac{\partial^{n-k} \mathcal{P}}{\partial \lambda^{n-k}} \\
    &= \sum_{k = 0}^n \frac{n!}{k!(n-k)!} \frac{\partial^k B_{n}}{\partial \lambda^k} B_{n-k} \mathcal{P}.
\end{align}
\end{subequations}

Then, integrating the equation over all the trajectories on both sides and assuming that the integral and the derivative are commutative. As a result, we have:
\begin{equation}
    \operatorname{Var}[B_n] = \langle B_n^2 \rangle = -\sum_{k=1}^n \frac{n!}{k!(n-k)!} \left\langle B_{n-k} \frac{\partial^k B_n}{\partial\lambda^k} \right\rangle.
\end{equation}

\section{An Example for $\operatorname{Var}[B_{n+1}] < \operatorname{Var}[B_n]$} \label{SIsec: example}

Consider a Bernoulli distribution:
\begin{equation}
    P(x; \lambda) = \begin{cases}
        \lambda, & x = 0; \\
        1 - \lambda, & x = 1,
    \end{cases}
\end{equation}
where $\lambda \in (0, 1)$ is the controlled parameter. The conjugate variables $\Lambda$ and $B_2$ are

\begin{subequations}
\begin{align}
    \Lambda(\lambda) = 
    \begin{cases}
        \frac{\partial\ln\lambda}{\partial\lambda} = \frac{1}{\lambda}, & x = 0; \\
        \frac{\partial\ln(1-\lambda)}{\partial\lambda} = -\frac{1}{1-\lambda}, & x = 1,
    \end{cases} \\
    \text{and} \quad
    B_2(\lambda) = 
    \begin{cases}
        \frac{1}{\lambda^2} - \frac{1}{\lambda^2} = 0, & x = 0; \\
        \frac{1}{(1 - \lambda)^2} - \frac{1}{(1 - \lambda)^2} = 0, & x = 1.
    \end{cases}
\end{align}
\end{subequations}
In this case, we have
\begin{align}
    \operatorname{Var}[\Lambda] &= \langle \Lambda^2 \rangle \\
    &= \lambda \times \left(\frac{1}{\lambda}\right)^2 + (1-\lambda) \times \left(-\frac{1}{1-\lambda}\right)^2 \\
    &= \frac{1}{\lambda} + \frac{1}{1-\lambda}, \\
    \operatorname{Var}[B_2] &= \langle B_2^2 \rangle = \lambda \times 0 + (1-\lambda) \times 0 = 0.
\end{align}
It shows that $\operatorname{Var}[B_2] < \operatorname{Var}[\Lambda]$. Therefore, $\operatorname{Var}[B_{n+1}] \ge \operatorname{Var}[B_n]$ is not generally true for all probability distributions.

\bibliography{manuscript}

\end{document}